# Strong reduction of $V^{4+}$ amount in vanadium oxide/hexadecylamine nanotubes by doping with $Co^{2+}$ and $Ni^{2+}$ ions: EPR and magnetic studies

M.E. Saleta [a], H.E. Troiani [a], S. Ribeiro Guevara [a], R.D. Sánchez [a*], M. Malta [b] and R.M. Torresi [c]

a) Centro Atómico Bariloche, CNEA, (8400) S. C. de Bariloche (RN), Argentina.

b) Depto. de Cs. Exatas e da Terra, Univ. do Estado da Bahia, Cabula Salvador (BA), Brazil.

c) Instituto de Química, Universidade de São Paulo, São Paulo (SP), Brazil.

**Abstract**

In this work we present a complete characterization and magnetic study of vanadium oxide/hexadecylamine nanotubes ($VO_x$/Hexa NT's) doped with $Co^{2+}$ and $Ni^{2+}$ ions. The morphology of the NT's has been characterized by Transmission Electron Microscopy (TEM) while the metallic elements have been quantified by Instrumental Neutron Activation Analysis (INAA) technique. The static and dynamic magnetic properties were studied collecting data of magnetization as a function of magnetic field and temperature and by Electron Paramagnetic Resonance (EPR). We observed that the incorporation of metallic ions ($Co^{2+}$, $S=3/2$ and $Ni^{2+}$, $S=1$) decrease notably the amount of $V^{4+}$ ions in the system, from 14-16% (non-doped case) to 2-4%, with respect to the total vanadium atoms into the tubular nanostructure, improving considerably their potential technological applications as Li-ion batteries cathodes.

## 1. Introduction

Oxide nanotubes (NT's) such as $SiO_2$, $TiO_2$, ZnO, $VO_x$ and so on, have been the subject of a very active research due to their distinctive physical and chemical properties, which have promising technological applications [1]. Among these compounds are the vanadium oxide ($VO_x$) NT's, in which the walls are constituted by alternated layers of $VO_x$ and surfactant. The surfactant or template helps to form the skeleton, which provides support and hardness to the NT's walls. Different templates have been used as primary monoamines ($C_nH_{2n+1}NH_2$) and $\alpha,\omega$-diamines ($H_2N(CH_2)_n NH_2$) [2, 3]. The layers of $VO_x$ are constituted by V ions that, in this case,

---

[*] Corresponding author: Tel.: +54 2944 445158; fax: +54 2944 445299: e-mail address: rodo@cab.cnea.gov.ar (R.D. Sánchez)

can be in two different oxidation states (4+, with a magnetic spin $S=1/2$ and 5+, the diamagnetic ions with $S=0$). The presence of $V^{4+}$ ions in the estructure is essential for the rolling-up of the $VO_x$ layer in order to form the NT [2]. More over, Vera-Robles and Campero [4] observed that the presence of both V species ($V^{4+}$ and $V^{5+}$) is fundamental for the scrolling process that occurs during hydrothermal treatment. They worked with vanadyl (IV) acetate and dodecylamine as precursor and obtained the V mixed valence by oxidizing the $V^{4+}$. In two previous works, Krusin-Elbaum [5] and Vavilova [6] studied $VO_x$/dodecylamine NT's and they reported a high percentage of $V^{4+}$ ions detected by magnetic experiments. Even more, half of these magnetic ions are antiferromagnetically coupled dimmers.

The oxide layer can be described using "V-O double layer" structure of the $BaV_7O_{16} \cdot nH_2O$ ($P4_2/m$) [5]. The structure is shown in Fig. 1. In the $V_7O_{16}$ structure both V(1) and V(2) sites are coordinated with six oxygen atoms ($VO_6$-octahedral). These octahedra form the vanadium oxide double layer. Between these layers, the remaining vanadium atoms are located in the V(3) sites and they are coordinated with four oxygen atoms forming tetrahedrons [7].

The effect of cation exchange in $VO_x$/surfactant NT's was initially studied by Nesper's group [8] and subsequently by other authors [9,10]. They reported that some transition metals, alkali metals and alkali-earths can be exchanged with good preservation of the tubular shape. Azambre and Hudson have also synthesized copper nanoparticles (NP's) within $VO_x$/dodecylamine NT's [11] and our group has studied nanocomposites constituted by $FeO_y$ NP's and $VO_x$/hexadecylamine NT's [12].

The $VO_x$ NT's present several potential technological applications, such as sensing elements [13], catalysis [14] and Li-ion batteries cathodes [15]. In the last case, it was found that the amount of $V^{4+}$ in the structure is critical as it affects and reduce notably the performance of the cathode.

In this context, we present a complete structural, magnetic and Electron Paramagnetic Resonance (EPR) study on $VO_x$/hexadecylamine nanotubes in which magnetic Co or Ni ions are intercalated in the structure. Our experiments confirm that the incorporation of $Co^{2+}$ and $Ni^{2+}$ ions into the amine layers reduces notably the amount of the isolated and paramagnetic $V^{4+}$ ions.

**2. Experimental**

The $VO_x$/Hexa NT's were synthesized in two steps. In the first one, 3.09g of crystalline $V_2O_5$ (Aldrich) were slowly added in a beaker containing an ethanolic solution (5.7mL) of hexadecylamine 90% (4.55g, Aldrich). The mixture was then maintained under magnetic stirring during two hours and, after that, 16.7mL of deionized water were added and the material was aged during 48h. The second step consists in a hydrothermal digestion of the gel during 7 days at 180°C in order to form the non-doped NT's. Afterwards, to obtain the doped NT's, the $VO_x$/Hexa nanostructures were introduced into a solution if ethanol/water (4:1 v/v) with the metallic ions, where a fraction of the amines were exchanged [8, 16]. We used $CoCl_2 \cdot 6H_2O$ (Merck) and $NiCl_2 \cdot 6H_2O$ (Merck) in each solution, respectively, in a molar ratio excess of 4/1 (metallic cation/$V_2O_5$-hexadecylamine). The solution was stirred for 3 hours and filtered, and the solid black product was rinsed and dried under vacuum.

The concentration of V and Co in the NT's was determined by Instrumental Neutron Activation Analysis (INAA). The samples were irradiated in the RA-6 research nuclear reactor, located at Centro Atómico Bariloche (Bariloche Atomic Center - Argentina), in a thermalized neutron flux ($\phi_{th} \cong 8 \times 10^{11}$ n.cm$^{-2}$.s$^{-1}$). Vanadium and cobalt concentrations were determined by comparison with high purity metallic standard materials (V: Johnson Matthey, 99.7% purity, and Co: R/X Reactor Experiments, 99.95% purity). Vanadium was measured by evaluating the 1.43 MeV emission of the $^{52}$V isotope (half-life of 3.743 minutes), while cobalt concentrations were determined by the 1.17 and 1.33 MeV emissions of $^{60}$Co (half-life of 1925.28 days). Gamma-ray spectra were collected after appropriate decaying times regarding activation products half lives. In the case of Co, the decay time was adjusted to allow complete decay of $^{60}$Co$^m$ (half-life of 10.467 minutes) to the ground state for the samples and standards. Due to the long half-life of the $^{58}$Ni(n,p) $^{58}$Co reaction (half-life of 70.86 days) we did not quantify the amount of Ni with this technique.

The tubular shape of the products was confirmed by Transmission Electron Microscopy (TEM) and the local composition was characterized by EDS. Both studies were performed in a CM 200 Philips microscope (LaB$_6$ cathode, 200 keV). The X-Ray Diffraction (XRD) patterns were acquired in a Philips PW 1700 diffractometer using an aluminum sample holder and CuK$_\alpha$ radiation (1.54186 Å).

EPR spectra were collected with a Bruker ESP-300 spectrometer, operating at X-band (9.5 GHz) varying the temperature between 100K and 300K. Two additional spectra at K-band (24 GHz) and Q-band (35 GHz) were taken at room temperature to complete the energy levels involved in the magnetic transitions.

The static magnetic characterization of the samples was performed in a commercial superconducting quantum interference device magnetometer (Quantum Design MPMS-5S and MPMS-XL) with fields up to 70kOe. The measurements as a function of temperature were performed in the 5K-360K range.

## 3. Results

### 3.1 Morphology and composition

In order to understand the shape formation mechanism of the NT's, we show in Fig. 2a and 2b TEM images of tubes that have not completely closed. Schematic representations of the rolling-up of planes are presented in Fig. 2c and 2d. These photographs confirm the previous rolling hypothesis mentioned by Nesper and Muhr [17].

A typical TEM image of non-doped NT is presented in Fig. 3a. The dark and bright lines are associated to the alternated layers of $VO_x$ and Hexa showing the multiwall characteristics of these NT's, present in all the studied samples. Comparing the doped and the non-doped NT's, we observe an increment of defects in the doped samples, like loss of definition and waving surface of the walls. These defects can be a consequence of the doping process when the Hexa was partially exchanged by the metallic ions. In the middle picture (Fig. 3b) the EDS spectra of the samples is shown. In the Co and Ni doped NT's, an extra peak can be clearly detected due to the presence of each doping element. The V and Co amounts, determined by INAA, are presented in Table 1. The atomic mol ratio quantification between Co and V using both techniques, local (EDS) and global (INAA), are in good agreement, yielding the value Co/V = 0.14. With the EDS technique we obtain 0.16 for the Ni/V ratio.

The XRD patterns are presented in Fig. 3c; we can considerer two regions in the patterns: the low angle (LA, $1.5º \leq 2\theta \leq 15º$) and high angle (HA, $15º \leq 2\theta$) region. The LA data have the reflections indexed as *00l*, which provide information about the interlamellar distance (*d*). On the other hand the HA measurements have the *hk0* reflections, which provide the $VO_x$ cell parameters; these last peaks have lower intensity than the LA peaks. In order to index the cell we assumed a planar square lattice [2]. The interlamellar distances decrease with doping, which is in agreement with the exchange of Hexa by Ni or Co ions. The values of *d* are presented in Table 1. The *hk0* reflections do not vary with the incorporation of Ni and Co, which indicates that the samples have practically the same cell parameters. This result is an indication that the metallic ions are not incorporated to the $VO_x$ double layered structure. In this way, we assume that either the doped ions are incorporated to the structure or they replace the amines in the skeleton of the NT's. The calculated cell parameter is $a \approx 5.93(2)$Å for the three samples.

3.2. EPR measurements

The X-band EPR spectra measured at room temperature are shown in Fig 4. From the bottom to the upper plot, the resonance lines for the three studied samples correspond to: non-doped (only magnetic vanadium ions), Co-doped (magnetic vanadium and cobalt ions) and Ni-doped (magnetic vanadium and nickel ions).

The three X-band EPR spectra can be described by the following Hamiltonian:

$$\mathcal{H} = \mathcal{H}_{V^{4+}} + \mathcal{H}_{i=Co^{2+},Ni^{2+}}, \qquad (1)$$

where $\mathcal{H}_{V^{4+}}$ is the Hamiltonian given by the $V^{4+}$ ions and $\mathcal{H}_{i=Co^{2+},Ni^{2+}}$ is the contribution to the spin Hamiltonian of the transition metal doped ion ($i=Co^{2+}$ and $Ni^{2+}$). In the non-doped sample only the Hamiltonian of the $V^{4+}$ contributes to the resonance, while for the doped samples we have to include the extra term of metal dopants to describe the absorption.

The non-doped sample can be adjusted with a single line centered at $g \approx 1.96$ and that can be associated with the $V^{4+}$ ($S=1/2$) ions. However, when the spectra are measured at higher frequencies (K and Q bands), the lines become asymmetric (different heights between the maximum -or minimum- and the base line). This asymmetry means that the sample presents crystalline anisotropy. The Hamiltonian which describes this anisotropy is presented in Eq. 2. This expression also includes the hyperfine interaction between the electronic and nuclear spin of $^{51}V$ (abundance 99.76%, $I=7/2$) [18].

$$\mathcal{H}_{V^{4+}} = \mu_B[(g_\parallel . H_z . S_z) + g_\perp(H_x . S_x + H_y . S_y)] + [(A_\parallel . S_z . I_z) + A_\perp(S_x . I_x + S_y . I_y)] \quad (2)$$

where $\mu_B$ is the Bohr magneton, $S_j$ and $I_j$ ($j = x, y, z$) are the projections in the $j$-direction of the electronic spin and nuclear spin operator respectively; $g_\parallel$ and $g_\perp$ are the parallel and perpendicular (respect to magnetic field - $H$) components of the $g$-factor, and finally $A_\parallel$ and $A_\perp$ are the principal (parallel and perpendicular) components of the hyperfine tensor in magnetic field units – G.

In order o fit correctly the EPR spectra it was necessary to considerer a random distribution of crystalline orientations (powder distribution) and we only took into account the Zeeman term (first term of the Eq. 2). The experimental and theoretical fittings are presented in Fig. 5. The obtained $g$ components of the non-doped NT's are shown in Table 2. Now, when we subtracte the calculated powder line from the experimental resonance, a low intensity spectrum formed by equally spaced lines remains. These absorptions correspond to the hyperfine structure (hfs) produced by the interaction between the electronic and nuclear spins of $^{51}V$ (both terms of Eq. 2). Contrary to those vanadium ions that produce the subtracted powder line, the hfs is an indication that some $V^{4+}$ ions are practically isolated without interacting with other $V^{4+}$ ions. In the case of the V ions which contribute to the powder line, the hfs collapses into a single line (width 150G, X-band). This effect can be explained by the exchange interaction between near $V^{4+}$ ions (exchange-narrowing). This effect was studied by Deigen and co-workers in $Mn^{2+}$ ions [19] and by Wiench et al. in $V^{4+}$ ions [20].

In order to describe the $^{51}V^{4+}$ EPR in our experiments, the magnetic field positions of the hfs satellites lines can be expressed as [18, 21]:

$$H_\perp(m_I) = \left(\frac{h\nu}{g_\perp \mu_B}\right) - A_\perp m_I - \left(\frac{63}{4} - m_I^2\right)\left(A_\parallel^2 + A_\perp^2\right)\left(\frac{g_\perp \mu_B}{h\nu}\right) \quad (3a)$$

$$H_\parallel(m_I) = \left(\frac{h\nu}{g_\parallel \mu_B}\right) - A_\parallel m_I - \left(\frac{63}{4} - m_I^2\right) A_\perp^2 \left(\frac{g_\parallel \mu_B}{h\nu}\right) \quad (3b)$$

where $h$ is the Plank's constant, $\nu$ the microwave frequency, and $m_I$ is the nuclear magnetic quantum number ($m_I = \pm 7/2, \pm 5/2, \pm 3/2, \pm 1/2$). From the fit of the data, we obtain $A_\parallel$ and $A_\perp$, presented in Table 2 along with bibliographic results to compare [22].

Our EPR results in non-doped $VO_x$ NT's are completely different respect to those reported by Kweon and co-workers [23], where they fitted their data assuming two wide resonance lines to explain their experiment. One of them presents similar satellites as those observed by us. However, their second broader (~ 750 G) resonance line associated to the absorption of $V^{4+}$-$V^{4+}$ dimmers is not present in our case.

The doped samples present more complex spectra because, in addition to the $V^{4+}$ lines complexity, they also present the resonance lines provided by the magnetic doped ion.

Our first studied case is the EPR in X-band of Ni-doped NT's (Fig. 6), where two contributions are easily visible. These contributions are described by the $V^{4+}$ and $Ni^{2+}$ Hamiltonians ($\mathcal{H} = \mathcal{H}_{V^{4+}} + \mathcal{H}_{Ni^{2+}} = \mathcal{H}_{V^{4+}} + g_{Ni^{2+}} \overline{H}.\overline{S}$). In the spectrum both contributions are presented; the first is a signal with a superstructure centered at $g \approx 1.96$ with a well resolved hfs. This signal corresponds to the $V^{4+}$ ions. The other broad line (~ 1400 G) is well described by a Lorentzian lineshape centered at $g_{Ni^{2+}} \approx 2.22$ which is in agreement with the expected value for $Ni^{2+}$ paramagnetic ions [24]. In order to adjust the EPR parameters of the Ni contribution we fitted the spectrum keeping the width and center field as free parameters. The fitted EPR parameters for both contributions are presented in Table 2; the g's and A's factors of the $V^{4+}$ contribution were calculated using Eqs. 3.

The Co-doped vanadium oxide NT's are a more complex case because we can not explain the behavior with only two components for the two magnetic ions (V and Co). To make a very good description, we need to introduce for both ions the $g_\parallel$ and $g_\perp$ contributions (crystalline anisotropy). The Co EPR resonance can be calculated considering quantum transition probabilities between the energy levels, the crystal field effect on these levels and its angular distribution. To describe the asymmetric Co resonance, we use a powder line-shape to fit the spectrum for each employed microwave frequency.

To obtain the position of the resonance field for each angle, we have to solve the following Hamiltonian [25]:

$$\mathcal{H}_{Co^{2+}} = H_{eff} \mu_B [(g_\perp - g_\parallel) \sin\theta \cos\theta\, S_x + (g_\perp \sin^2\theta + g_\parallel \cos^2\theta) S_z] +$$
$$D\{(\cos^2\theta - \sin^2\theta)[S_z^2 - \frac{1}{3}S(S+1)]$$
$$+ \cos\theta \sin\theta (S_x S_z + S_z S_x) - \frac{1}{2}\sin^2\theta (S_x^2 - S_y^2)\} \qquad (4)$$

where $H_{eff} = H - H_0$ is the effective magnetic field with $H_0$ being an internal field and $H$ the external magnetic field; $D$ is the orthorhombic crystal field parameter and $\theta$ is the angle between the $z$ direction of the crystal and the external magnetic field ($\theta = 0°$, parallel and $\theta = 90°$, perpendicular).

Finally, the Co line ($Y(\theta, H_0)$) was calculated by adding each angular contribution weighted by the angle position ($\sin\theta$):

$$Y(\theta, H_0) = \sum_\theta \sum_{\substack{i \neq j \\ i,j=1,2,3,4}} I_{ij}(\theta, H_{ij}) f(\theta, H_{ij}, H_0) \sin\theta \qquad (5)$$

where $f(\theta, H_{ij}, H_0)$ is the Lorentzian line describing the absorption as a function of the polar angle ($\theta$), the resonance field ($H_{ij}$) and the internal field ($H_0$); $I_{ij}(\theta, H_{ij})$ is the intensity of each transition and angle, calculated from the eigenvectors of the Hamiltonian matrix. $H_{ij}$ is the magnetic field where the difference between two energies ($E_i$ and $E_j$) is equal to $h\nu/\mu_B$. The energy levels are the eigenvalues of the Hamiltonian matrix

In Fig. 7a and 7c we show the magnetic field dependence of the energy levels for the two extreme cases, $\theta = 0°$ and $\theta = 90°$. In Fig. 7b, the magnetic field dependence for intermediate polar angles is represented. The calculated intensities ($I_{ij}$) of the low field transitions (1-3, 1-4 and 2-4, see left vertical lines in Fig. 7) are negligible, while the high field lines contribution to the total resonance is much stronger. The envelope curves, for the three frequency bands, can be well fit with the proposed cobalt model (see solid line in Fig. 8a, 8b and 8c).

The superstructure of the $V^{4+}$ was described by both terms of $\mathcal{H}_{V^{4+}}$ (Eq. 2). We fit the position of the hfs resonances by Eqs. 3a and 3b. In the inset of Fig. 8a we show the experimental and calculated magnetic field position ($H_i$) for parallel and perpendicular hfs vanadium lines as a function of $m_I$. On the other hand, in the inset of Fig. 8b, we plot the residual experimental data after subtracting the model for the Co contribution (fitted by Eq. 5). The upper and lower bars in this inset indicate resonance fields of the perpendicular and parallel contribution of the hfs of the vanadium ions and we show how these bars coincide with the position of the resonances in the spectrum. The fitting parameters are shown in Table 2.

We remark that Co-doped NT's and Ni-doped NT's systems, after the subtraction of the doping ion resonance, show the same residual lines in amount and magnetic field positions which can be directly associated to the hfs contribution of $V^{4+}$ ions. Indeed, $^{58}$Ni (68.0% abundance) and $^{60}$Ni (23.2% abundance) with $I=0$ do not present hfs resonances. On the other hand, although $^{59}$Co has nuclear spin ($I=7/2$) and approximately a 100% of abundance, no hfs corresponding to these atoms was detected du to the exchange narrowing effect. These features constitute good evidence that the observed residual lines are only produced by the hfs $^{51}$V in both doped samples (Co and Ni). Also, we can assert that the observation of the well-resolved $V^{4+}$ hfs resonances indicates: *i*) the presence of isolated $V^{4+}$ magnetic ions and *ii*) that these doped samples have a considerable smaller $V^{4+}$ amount than the non-doped $VO_x$ NT's.

3.3. dc-magnetic characterization

The dc-magnetization as a function of temperature of non-doped and doped NT's measured with an applied field of 10kOe is presented in Fig. 9.

In a previous work we have described the magnetic behavior of Ni-$VO_x$ NT's as a paramagnet which follows the law: $\chi(T)=C/(T-\Theta)+ \chi_0$ (Curie-Weiss model that a temperature independent term), where $C$ is the Curie constant, $\Theta$ is the Curie's temperature, $\chi_0$ is a temperature independent contribution, which has been calculated following the procedure previously described in Ref. [26]. In the studied temperature range the samples did not present antiferromagnetic dimmers, as was reported by other authors [5, 6]. In the Ni and Co doped samples we assume that $C$ has two contributions: one from the paramagnetic doping species ($Ni^{2+}$ or $Co^{2+}$) and the other from to the $V^{4+}$ ions. As the experimental results were normalized by the moles of V, the units of $C$ are expressed in moles of V too. Them the Curie constant can be written as:

$$C = \sum_{i=V^{4+},Co^{2+},Ni^{2+}} \frac{N_A \cdot \mu_B^2 \cdot \langle g_i \rangle^2}{3k_B} S_i(S_i+1) \cdot F_{i/V} , \qquad (6)$$

where $k_B$ is the Boltzmann constant, $<g_i>$ is the mean value of the *g*-factor, $S_i$ is the spin of the *i-th* magnetic ion and; $F_{i/V}$ is the ratio *i*-mole/V-mole. From the Curie constant we calculated the $V^{4+}$ percentage in the non-doped NT's to be 16(2)%. Alternatively, for the doped case, the amount of $V^{4+}$ can not be calculated directly from $C$ and to obtain this we need to subtract first the contribution from Co and Ni. To calculate the Co and Ni contributions to $C$, we take into account our EPR results: the ions are present as $Co^{2+}$ ($3d^7$; $S=3/2$) and $Ni^{2+}$ ($3d^8$; $S=1$) and their

<g>-factors correspond to these in Table 2. Those reported considerations allow us to estimate the $V^{4+}$ amount in the NT's and the results are presented in Table 3.

To verify the amount of $V^{4+}$ obtained for the three samples from the temperature experiments we performed measurements varying the magnetic fields at low temperatures. The *M* vs. *H/T* curves do not follow a linear trend (Fig. 10) because our diluted paramagnetic ions are partially saturated by *H/T*. However, we obtained a very good fit with the model presented in Eq. 7. In all cases we have to take into account two terms: the linear contribution (second term in Eq. 7) and a Brillouin function [27] corresponding to $V^{4+}$ ions. When we consider the $Co^{2+}$ or $Ni^{2+}$ case, we need to add an extra Brillouin contribution for the transition metal doped ion. Them magnetization can be described by:

$$M(H/T) = \sum_{i=V^{4+},Co^{2+},Ni^{2+}} M_0^i . B_{S_i}(x_i) . F_{i/V} + \chi_0 . H \tag{7}$$

where $M_0^i = g_i S_i \mu_B$, $B_S(x_i)$ with $x_i = g_i \mu_B S_i H/(T.k_B)$ is the Brillouin function and its argument that describes the paramagnetic behavior of $i=V^{4+}$, $Co^{2+}$ or $Ni^{2+}$, as appropriate. The parameters obtained from the fitting of the *M* vs. *H* curves are presented in Table 3.

The two independent magnetic experiments (varying the temperature and varying the magnetic field) in the non-doped NT's are consistent. In both cases we obtained between 14 and 16% of quasi-free spins, which are responsible of the Curie-Weiss behavior observed in the magnetic susceptibility and magnetization. This result is also in good agreement with the vanadium percentage that occupies the V(3) sites, 1/7 of the total vanadium sites present in the unit cell. On the other hand, the V(3) are magnetically uncoupled because they are practically isolated from their vanadium neighbors. Both our quantitative experimental EPR and magnetization results of the quasi-free $V^{4+}$ ions and those reported by Krusin-Elbaum [5] and Vavilova [6] in similar $VO_x$ NT's, suggest that this percentage is a general characteristic of the $VO_x$ system and this not depend on the type of amine located between the oxide planes.

## 4. Conclusions

We observed in both the doped and non-doped samples that a fraction of the V ions that constitute the $VO_x$ NT's is in the $V^{4+}$ state. This fraction is reduced significantly in the Co and Ni doped samples which present sharp residual hyperfine $V^{4+}$ EPR resonances that are due to isolated $V^{4+}$ ions. In these samples a broad line is also detected, corresponding to the doped metal ion in each case. The broad Co-doped line is well described by a model which considers the powder resonance of $Co^{2+}$ ions with S=3/2 in an axial angular dependence of the cubic crystal field, perpendicular and parallel g-factors and Zeeman contribution. On the other hand, the broad Ni-doped line can be described by a single Lorentzian line of $Ni^{2+}$ ions with $<g>$=2.22.

The dc-magnetization studies (*M* vs. *H* curves), confirm the presence of *S*=3/2 coming from $Co^{2+}$ ions (in Co-doped NT's) and *S*=1 corresponding to $Ni^{2+}$ (in the case of Ni-doped samples). In addition the dc-magnetization allowed us to quantify the percentage of paramagnetic $V^{4+}$ ions present in both samples. The $VO_x$/Hexa NT's have a fraction of $V^{4+}$ between 14-16% while in the Co-doped sample it is reduced to 5-2% and close to 4-2% for Ni-doped NT's. This result is in good agreement with the observable $^{51}V$ hfs in both doped NT's.

Finally, we have presented an alternative method for doping vanadium oxide/hexadecylamine nanotubes with $Co^{2+}$ and $Ni^{2+}$, which allows to reduce the concentration of $V^{4+}$ ions keeping the tubular structure. The presence of these dopants can lead to an interesting modification in the nanotubes in order to minimize the $V^{4+}$ amount which is known to reduce the performance of this material when it is used as a cathode in Li batteries.

**Acknowledgements.** MES acknowledges to CONICET for the studentship. HET and RDS are members of CONICET. This work was partially funded by the following projects: in Argentina by U.N. Cuyo 06/C203; ANPCyT (PICT-2004 21372, PAV and RN3M); in Brazil by FAPESP (Proc. 03/10015-3).

Tables

Table 1: percentage of V and Co atoms measured by INAA and metal/V ratio calculated from EDS data.

|  | % V (INAA) | % metal (INAA) | metal/V ratio (EDS) | $d$ distance (XRD) |
|---|---|---|---|---|
| Non-doped $VO_x$/Hexa | 37.4 (9) | - - - | - - - | 2.1 |
| Co-doped $VO_x$/Hexa | 30.9 (5) | 4.2 (3) | 0.14 | 1.2 |
| Ni-doped $VO_x$/Hexa (a) | 32.6 (7) | - - - | 0.16 | 1.3 |

(a) previously reported in [11]

Table 2: Principal values of the hfs EPR spectrum measured at room temperature. The values for a $V_2O_5$ single crystal and amorphous were reported in Ref. 23.

|  | Metal | | | $V^{4+}$ | | | | |
|---|---|---|---|---|---|---|---|---|
|  | $g_{//}$ | $g_\perp$ | $<g>$ | $G_{//}$ | $g_\perp$ | $<g>$ | $A_{//}$ [G] | $A_\perp$ [G] |
| Non-doped $VO_x$/Hexa |  |  |  | 1.93 (1) | 1.96 (1) | 1.95 | 188 (5) | 92 (2) |
| Co-doped $VO_x$/Hexa * | 1.98 (2) | 1.95(2) | 1.96 | 1.930(5) | 1.980(5) | 1.963 | 185 (5) | 74(1) |
| Ni-doped $VO_x$/Hexa |  |  | 2.22 | 1.930(5) | 1.980(5) | 1.963 | 192(5) | 79(4) |
| $V_2O_5$ single crystal [23] |  |  |  | 1.923 | 1.986 | 1.965 | 187 | 67 |
| $V_2O_5$ amorphous [23] |  |  |  | 1.926 | 1.984 | 1.965 | 211 | 79 |

* $H_0$ = -483 (5)G and $D$= 433 (20) G.

Table 3: Fitting parameters obtained from the magnetic $\chi$ vs. $T$ and $M$ vs $H/T$ curves.

|  | Magnetic susceptibility curve [a] | | | $M$ vs $H$ curve | |
|---|---|---|---|---|---|
|  | $C$ [emu.K/V mole Oe] | $\theta$ [K] | %$V^{4+}$ | %$V^{4+}$ | Metal/V ($R$) |
| Non-doped $VO_x$/Hexa | 0.0606 (2) | -3.90 (4) | 16 (2) | 14 (1) | - - - |
| Co-doped $VO_x$/Hexa | 0.269 (9) | 2.26 (6) | 4.6 (4) | 2 (1) | 0.12 (1) |
| Ni-doped $VO_x$/Hexa | 0.2128 | -1.74 (5) | 4 (2) | 2 (1) | 0.13 (1) |

(a) we assume the Co/V ratio measured by INAA and EDS (0.14 for Co and 0.16 for Ni). In a previous work [11] we have already presented preliminary results of the magnetic characterization for Ni-doped NT's.

**Figures Captions**

Figure 1: Structure of the $VO_x$ layer, the V(1) and V(2) sites have octahedral coordination, while V(3) sites are in the center of a tetrahedral of oxygen ions. (a) Detail of the V(1,2)-O double layer. Between them are the V(3) sites, where vanadium atoms are less magnetically coupled with other vanadium atoms. (b) The same schematic array of atoms showing the polyhedra structure that forms the environment of oxygen around the vanadium atoms.

Figure 2: TEM micrographs of an uncompleted formed (a) Ni-doped NT and (b) non-doped NT. (c) and (d) schematic representation of the rolling-up process of the NT's.

Figure 3: (a) TEM image of a non-doped NT. (b) EDS spectra of non-doped and doped NT's, the Cu lines as produced by the sample holder. (c) Low angle (left) and high angle regions (right) XRD patterns for the non-doped and doped samples

Figure 4: X-band EPR spectra of $VO_x$/Hexa, Co-$VO_x$/Hexa and Ni-$VO_x$/Hexa NT's recorded at room temperature.

Figure 5: EPR spectra of $VO_x$/Hexa NT's recorded at room temperature, the experimental data were fitted using a powder line shape for ions with $S=1/2$ ($V^{4+}$). (a) X-band (9.5GHz), (b) K-band (24GHz) and (c) Q-band (35GHz).

Figure 6: (Color online) EPR spectrum of Ni-doped NT's collected at X-band and room temperature. The Ni signal was fitted with a Lorentzian function to describe the absorption curve; we also indicate in the upper and lower parts of the graph the positions of the V hfs lines (small vertical lines).

Figure 7: (a) Energy levels along $z$ ($\theta = 0°$). Calculated transitions at Q-band are indicated by vertical lines. (b) Angular dependence of the resonance magnetic field. (c) Energy levels corresponding to the $xy$ plane ($\theta = 90°$) configuration, vertical lines indicate the resonances at Q-band.

Figure 8: (color online) Room temperature EPR spectra of Co-doped NT's collected at: (a) X-band, (b) K-band and (c) Q-band. The noisy (black) curve is the experimental data and the smooth solid (red) line is the calculated spectrum for the Co contribution assuming a powder distribution. The inset in (a) shows hfs lines of the V contribution fitted with Eqs. 1a and 1b. Inset in (b) is a detail of the hfs of the V ions.

Figure 9: (Color online) Magnetic susceptibility as a function of temperature (solid circles: non-doped $VO_x$/Hexa NT's; open circles: Co-doped NT's; stars: Ni-doped NT's). Inset: $(\chi-\chi_0)^{-1}$ vs. temperature. $\chi_0$ involves all diamagnetic contributions and its calculation is mentioned in the text.

Figure 10: (color online) ($M_{molar} - \chi_0.H$) as a function of H, for (a) non-doped $VO_x$/Hexa NT's, (b) Co-doped NT's, (c) Ni-doped NT's. In all the samples the solid lines correspond to the fit using Eq. 7 (see text). For both doped samples we also plot the contribution from each magnetic ions.

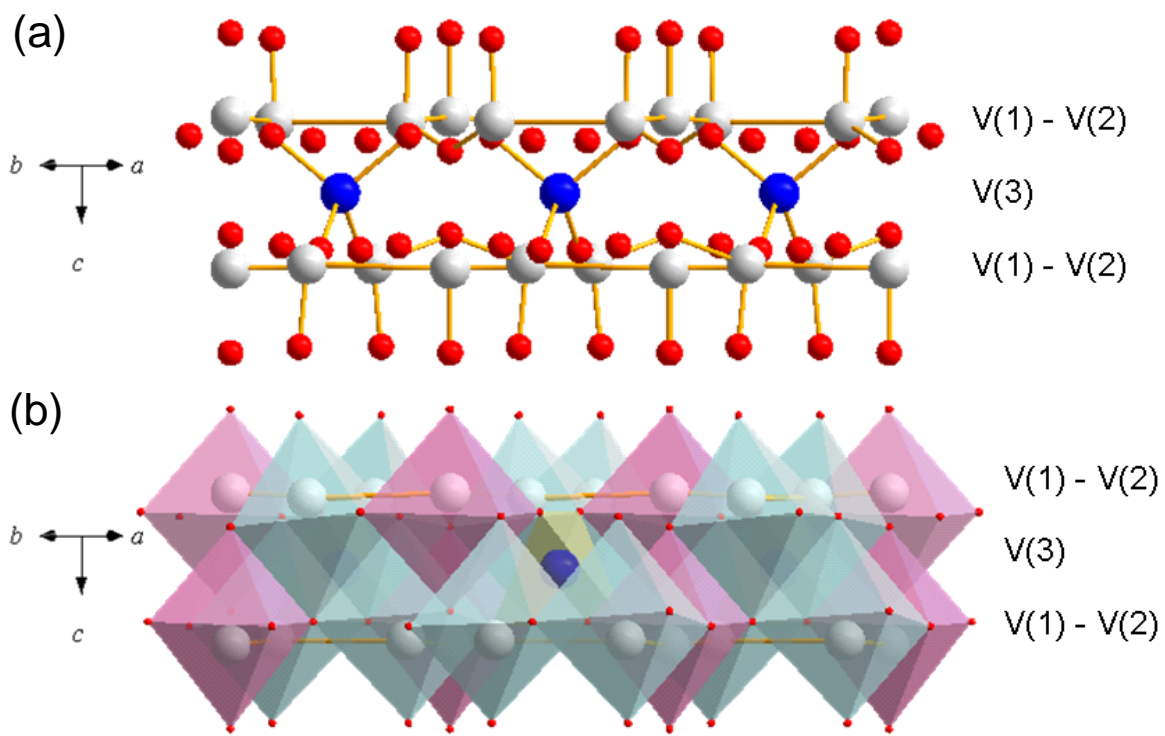

Figure 1

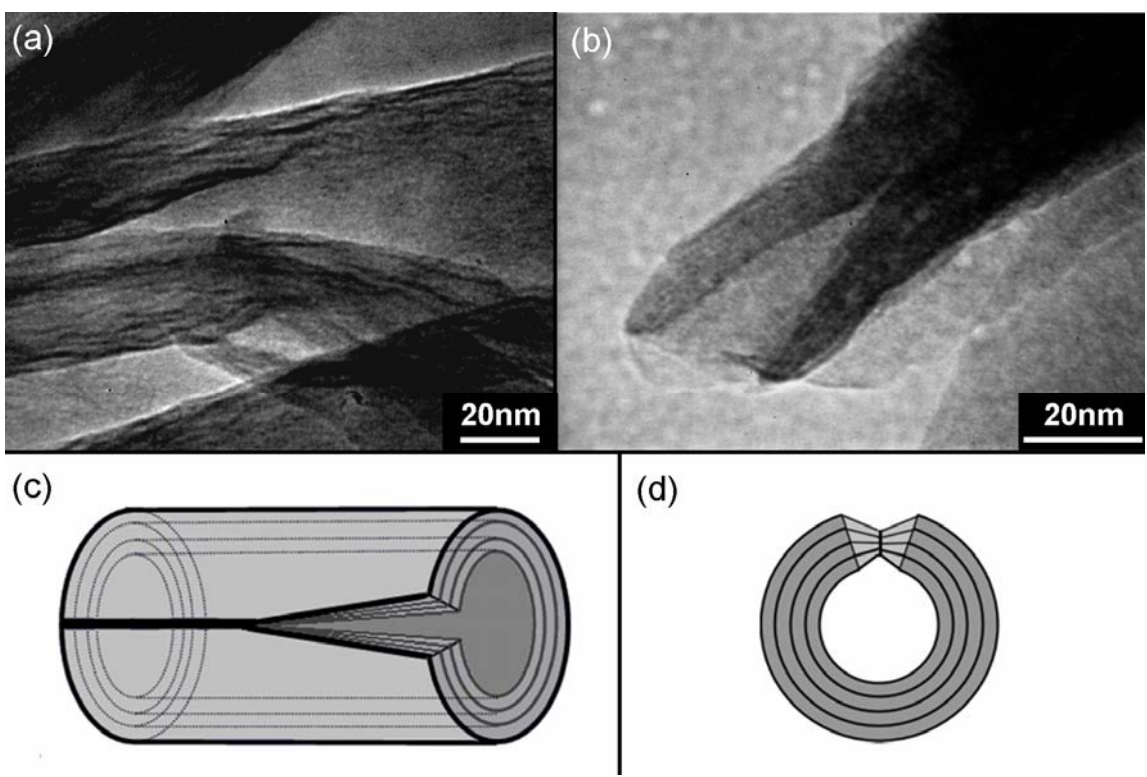

Figure 2

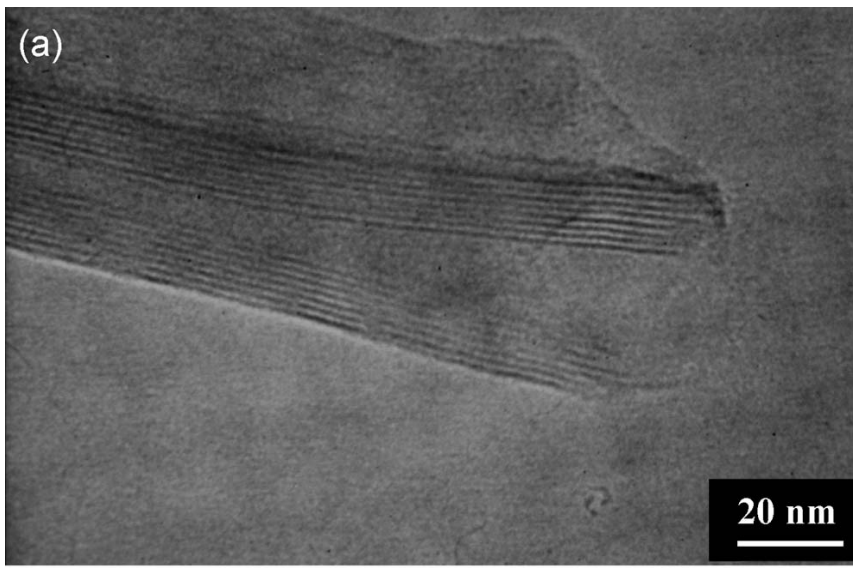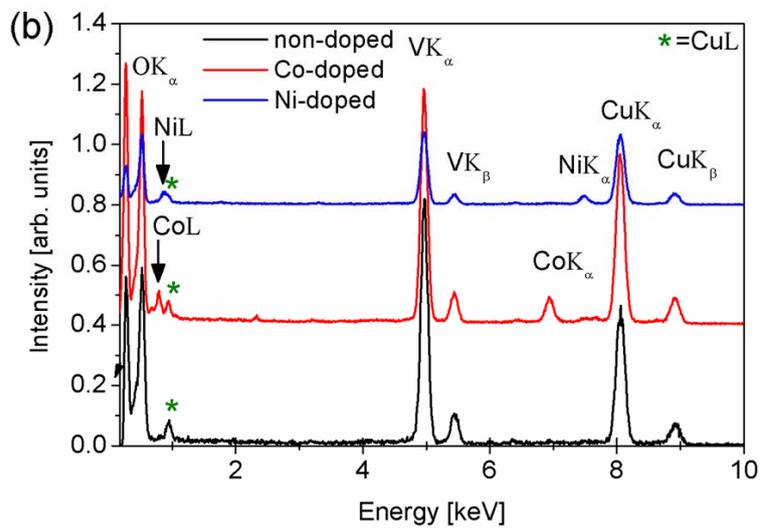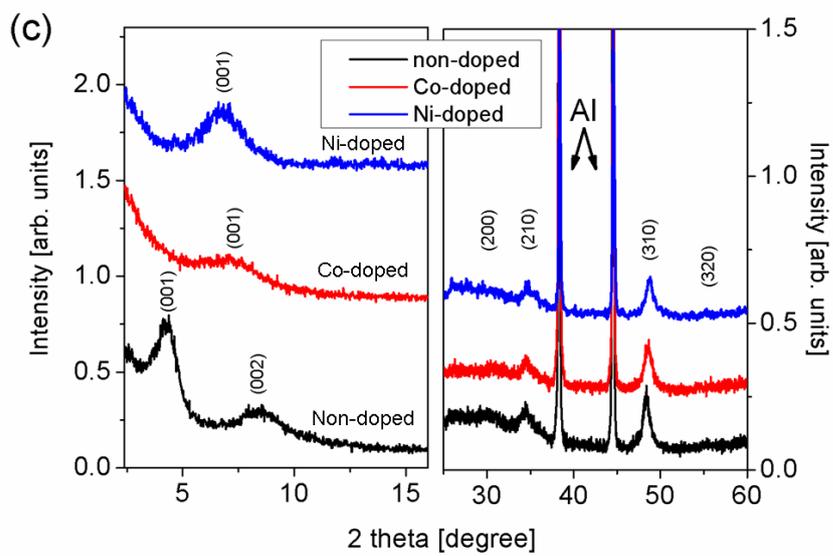

Figure 3

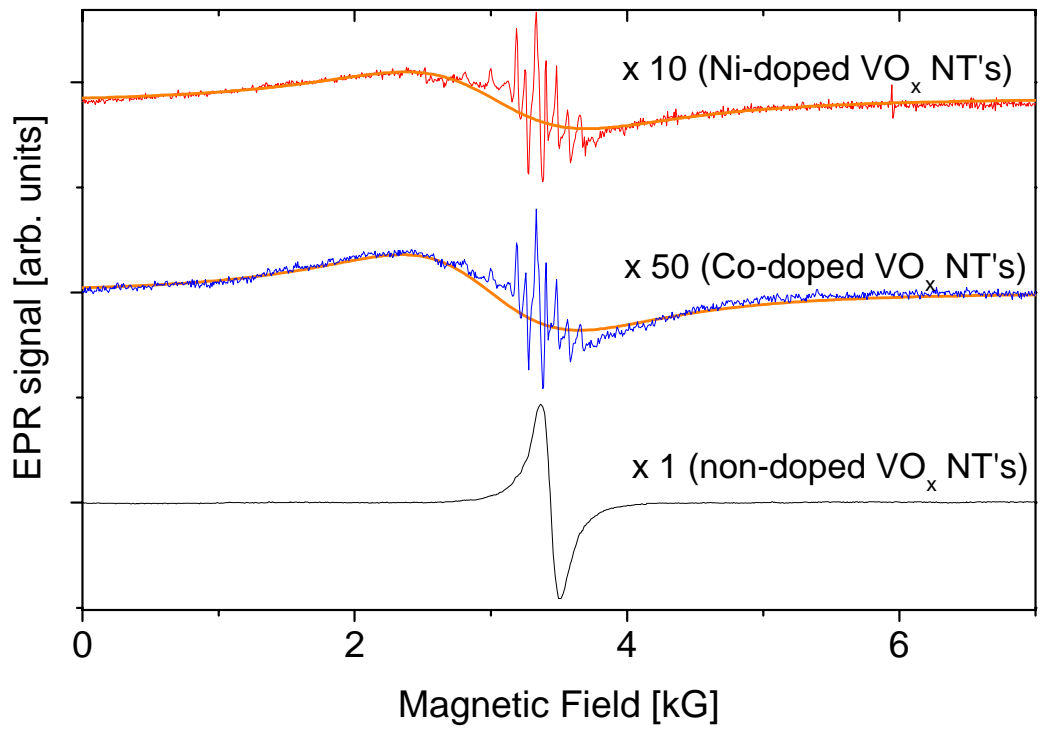

Figure 4

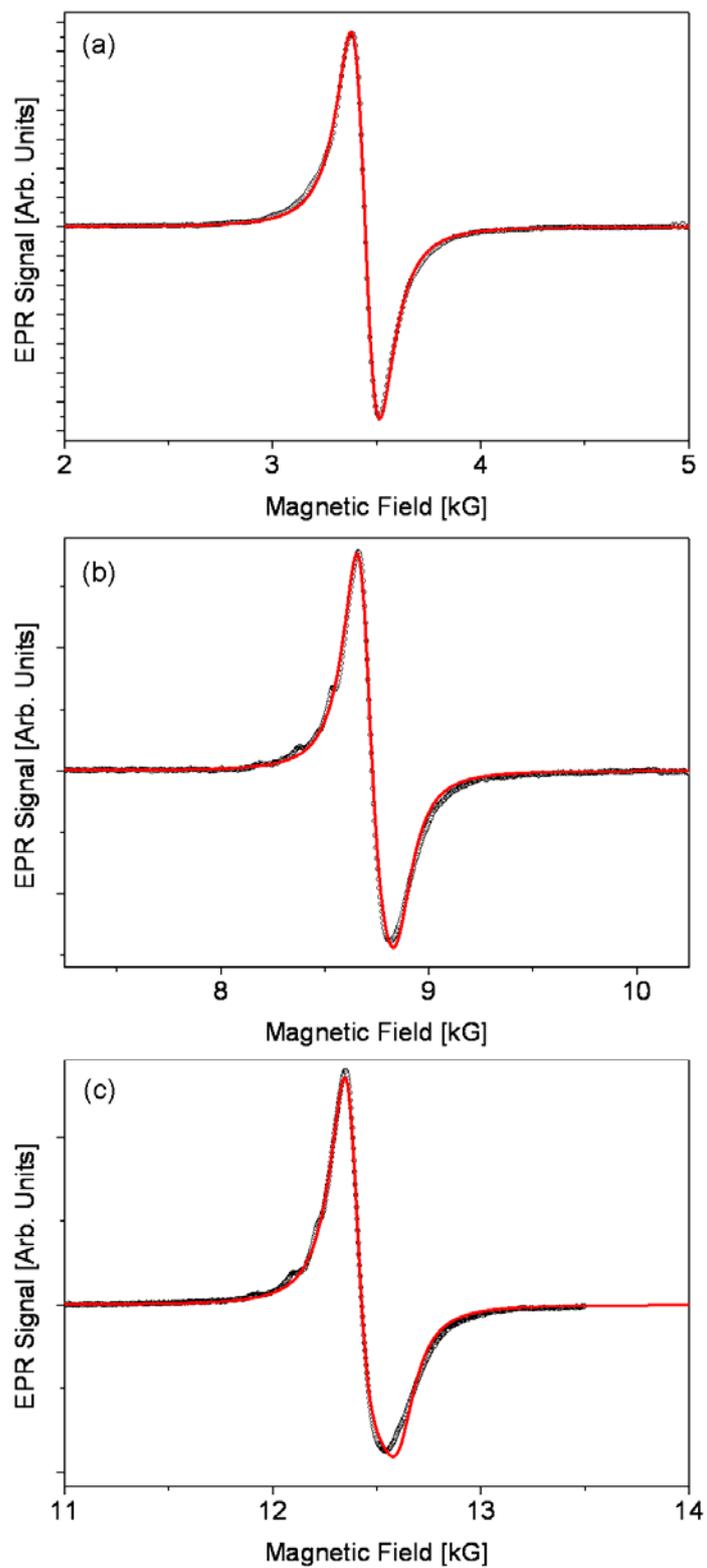

Figure 5

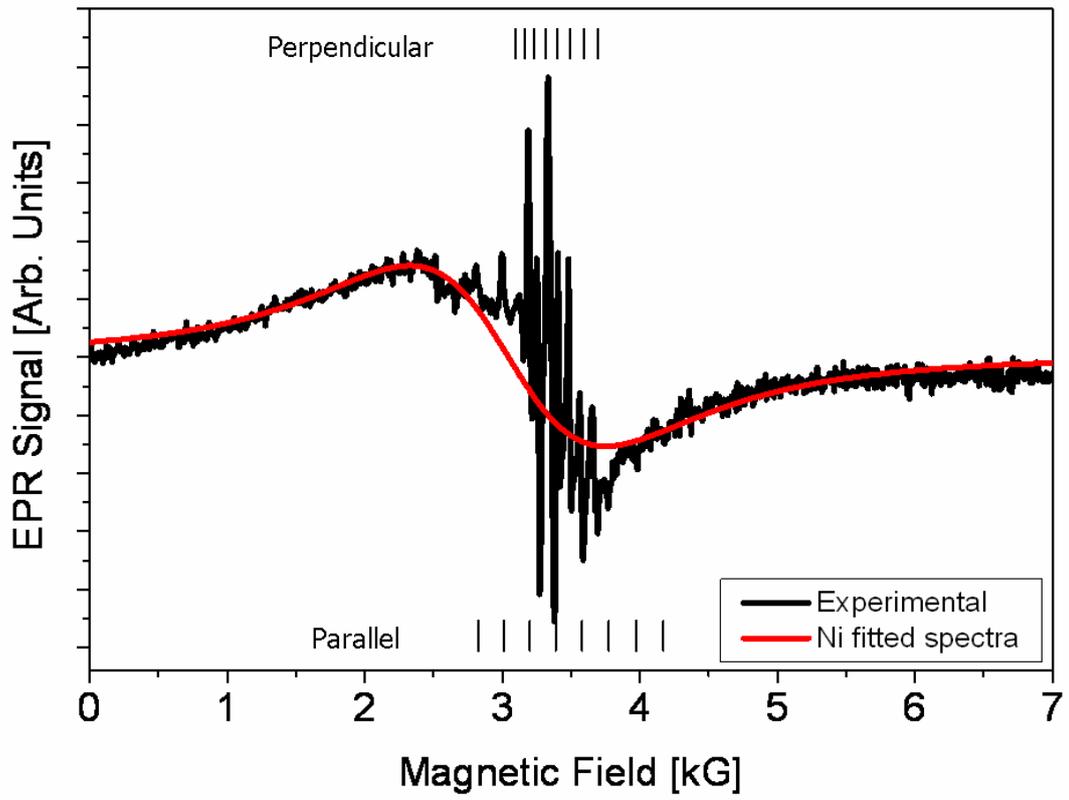

Figure 6

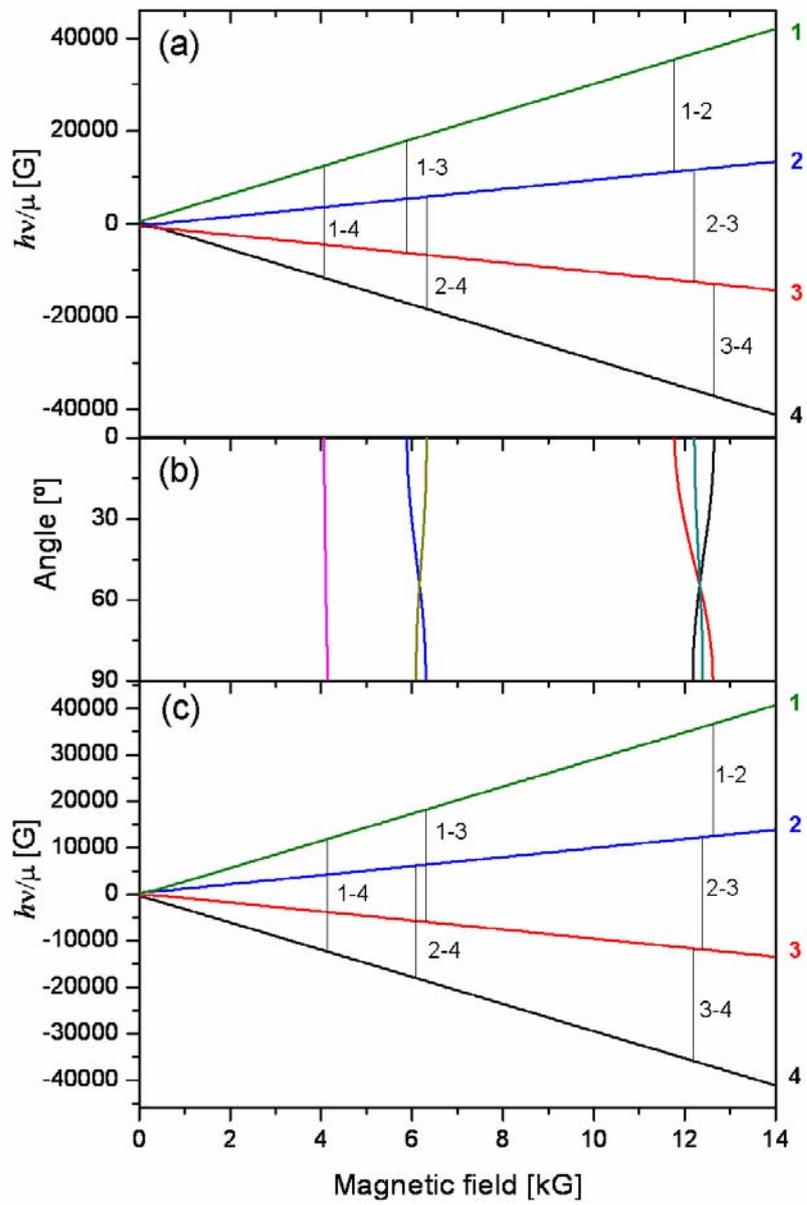

Figure 7

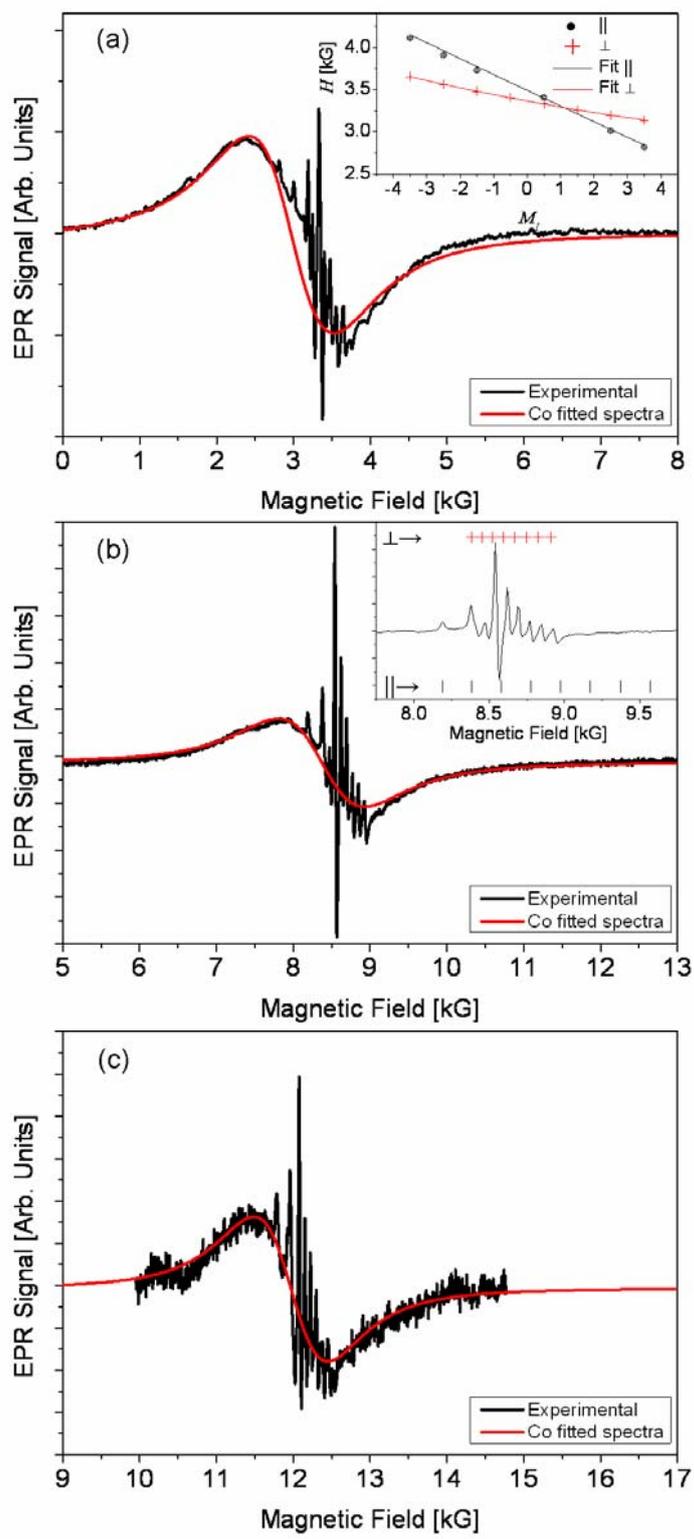

Figure 8

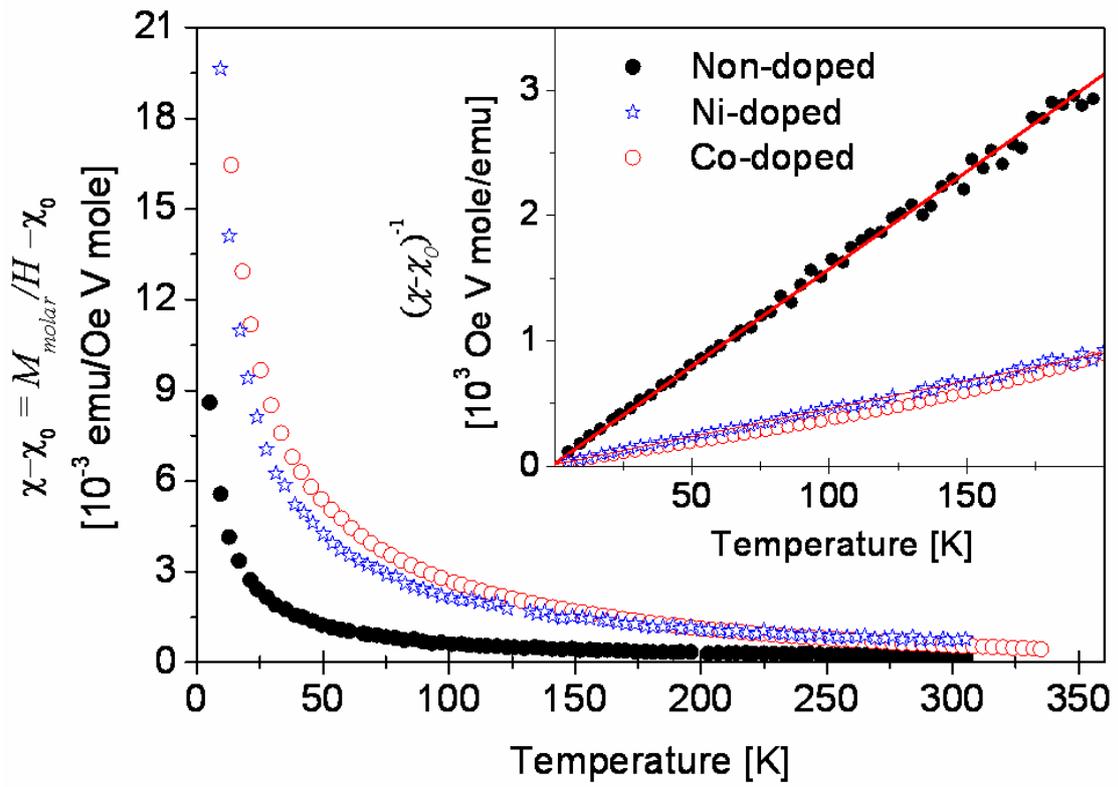

Figure 9

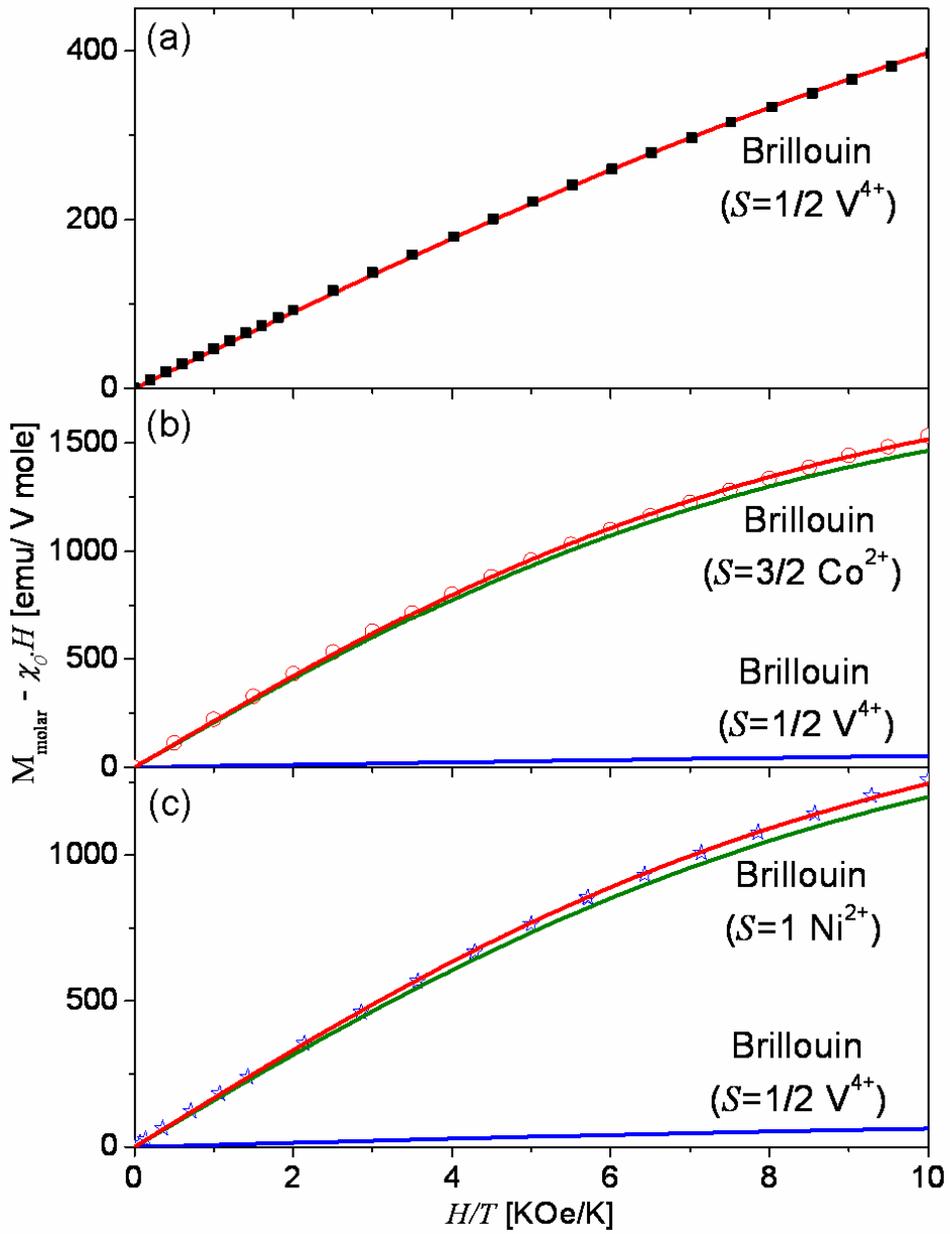

Figure 10